\documentclass[conference]{IEEEtran}
\IEEEoverridecommandlockouts
\usepackage[caption=false,font=footnotesize]{subfig}
\usepackage{amsmath,amssymb,amsfonts}
\usepackage{framed}
\usepackage{graphicx}
\usepackage{textcomp}
\usepackage{cite}
\usepackage{booktabs}
\usepackage{array}
\usepackage{float}
\usepackage{placeins}
\usepackage{stfloats}
\usepackage[font=small]{caption}
\usepackage{balance}
\usepackage{enumitem}
\usepackage[colorlinks,bookmarksopen,bookmarksnumbered,citecolor=blue,linkcolor=blue,urlcolor=blue]{hyperref}
\usepackage[capitalize]{cleveref}
\crefname{figure}{Fig.}{Figs.}
\Crefname{figure}{Fig.}{Figs.}
\crefname{subfigure}{Fig.}{Figs.}
\Crefname{subfigure}{Fig.}{Figs.}

\usepackage[
  paper=letterpaper,
  top=0.73in,
  bottom=1.01in,
  left=0.64in,
  right=0.64in,
  columnsep=0.25in
]{geometry}

\hypersetup{pdftitle={A User-Triggered UAV Dispatching System for Mountain Search},pdfauthor={},pdfsubject={},pdfkeywords={}}

\begin{document}
\raggedbottom
\brokenpenalty=10000

\title{A User-Triggered UAV Dispatching System for Precise and Timely Mountain Search Missions}

\author{
\IEEEauthorblockN{%
Mingyang Wang,
Yi Hong,
Tungchak Lee,
Ashwin Sundar, 
Kevin Hung,
Qubeijian Wang, and
Yalin Liu\textsuperscript{\dag~}
}
\IEEEauthorblockA{School of Science and Technology, Hong Kong Metropolitan University, Hong Kong SAR, China}
\IEEEauthorblockA{The School of Cybersecurity, Northwestern Polytechnical University, China}
\IEEEauthorblockA{Emails: \{s1465203, s1269615, s1313947\}@live.hkmu.edu.hk, \\ \{miwang, khung, ylliu\}@hkmu.edu.hk, qubeijian.wang@nwpu.edu.cn}

\thanks{\dag~Corresponding author.}
}


\maketitle

\begin{abstract}
Smartphones carried by hikers routinely record routes, locations, and requests for assistance, and these records can provide precise clues for mountain search. However, they are rarely transformed in a direct and timely way into unmanned aerial vehicle (UAV) missions that support search and rescue. This paper presents a user-triggered UAV dispatching system that integrates mobile information collection, ground station mission management, and UAV search into a single workflow. In our system, a mobile application is designed to continuously track user records and synchronize these records to the cloud database, thereby initiating all subsequent search events and UAV-dispatch missions. A control platform is developed to retrieve records from the cloud database for operator verification and mission scheduling. A UAV prototype is implemented to execute the approved mission, follow waypoints, report status, relay phone records when available, and record downward-facing video for later inspection. 
Simulations are conducted for UAV-dispatching missions using a terrain-following flight altitude of $80$~m above ground level and demonstrate successful route generation, ordered waypoint traversal, and return to the launch site for landing of the UAV dispatching. Local software tests verified timeout detection with a $1$s threshold and phone-record transfer from the mobile application through the UAV to the ground station. Outdoor trials showed that the UAV could fly autonomously and capture downward-facing imagery in which a person, a nearby footpath, and surrounding vegetation were clearly discernible, providing visual cues for operator assessment and further inspection of the search area. 
\end{abstract}

\section{Introduction}
\label{sec:intro}
Mountain search and rescue is time sensitive~\cite{adams_time_critical}. In Hong Kong, mountain rescue callouts increased from 215 in 2019 to 602 in 2020 and 951 in 2021~\cite{hkgov_mountain_rescue_2021,fsd_review2021}. Uneven terrain and sparse infrastructure also degrade mobile network coverage in country parks~\cite{ofca_coverage}. In this setting, routes, location records, and assistance requests left by hikers are often the most direct clues for organizing searches. Existing services already use hikers’ phones to support emergency response. Hikers can report nearby distance post numbers to identify their location~\cite{afcd_distpost}. The Hong Kong Police Force HKSOS application provides trip and location information to the 999 Reporting Centre during assistance requests, and can notify the centre and emergency contacts of prolonged journeys, route deviations, and abnormal behavior~\cite{police_hksos}. These services improve situational awareness but still require operators to interpret disparate records and to plan any aerial search manually. There is no direct pipeline from hiker side information to a search mission that can be dispatched under operator supervision. 

With high maneuverability, the emerging unmanned aerial vehicles (UAVs) are widely used to support victim search missions in diverse, complex terrains. Related studies span wireless device localization~\cite{calabro_transponder_2024,blaga_cellular_localization_2025}, user localization with integrated sensing and communication~\cite{moro_isac_2024}, wilderness search using phone Wi-Fi signals~\cite{wi2sar_2026}, cooperation among UAVs~\cite{lanca_multi_uav_2026}, and person detection in visible~\cite{dumencic_wilderness_2025} and infrared imagery~\cite{kim_cloud_ai_2025}. At the system level, Fly4SmartCity combines a cloud platform and UAVs for urban emergency services~\cite{ermacora_fly4smartcity_2016}, and AUSPEX integrates mission planning and execution with human supervision~\cite{doschl_auspex_2025}. These efforts focus on aerial sensing, localization, and general-purpose mission management. They do not address a workflow that starts from user-initiated hiking records and assistance requests, nor do they target mountain environments with intermittent connectivity and the need to interpret partial phone traces.

In this paper, we propose a user-triggered UAV dispatching system that uses hikers' smartphones to connect user information collection, mission dispatch, and UAV search. After the operator verifies the need for a search, the system generates a flight route from the available user information for review and dispatch, and the UAV then executes the confirmed mission automatically. The contributions of this work are threefold. First, we design a complete workflow that connects hiker-side information collection, cloud-based event management, and UAV search, enabling user-triggered dispatch of aerial missions for mountain search missions. Second, we develop a mobile application with Event Booking, Quick Start, and SOS modes, together with rule-based handling of location update timeouts and anomaly alerts that drive the creation of search events and UAV routes. Third, we implement a prototype and evaluate it through software tests that validate timeout and record transfer logic, and outdoor field trials that demonstrate autonomous UAV flight and ground image collection suitable for operator review. Overall, this integrated approach shows how information already available on hikers’ smartphones can be turned into actionable UAV missions, reducing reaction time and supporting more precise searches in mountain environments.

\section{System Design}
\label{sec:design}
\begin{figure*}[!t]
    \centering
    \includegraphics[width=0.9\textwidth]{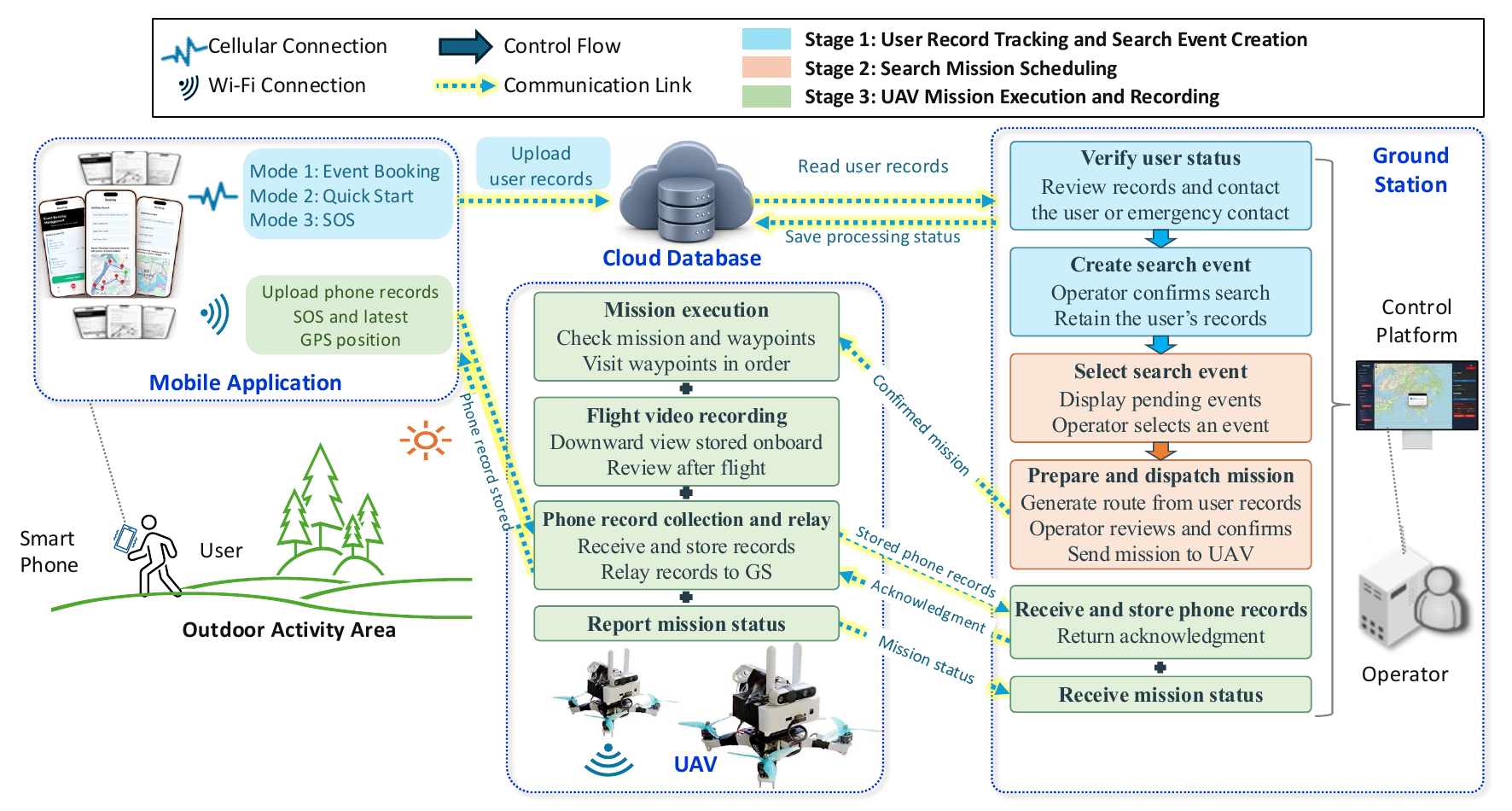}
    \caption{The three-stage workflow of the user-triggered UAV dispatching system. } 
    \label{fig:system_architecture}
\end{figure*}

By integrating user route data with the UAV's trajectory plan, we design a ``user-triggered UAV dispatching system'' to support precise and timely responses in mountain search missions. As illustrated in \Cref{fig:system_architecture}, the system comprises a mobile application on the user's device (i.e., smartphone), a cloud database, a control platform in the ground station, and a UAV for search missions. The mobile application continuously tracks user records and synchronizes these records to the cloud database, thereby initiating all subsequent search events and UAV-dispatch missions. The control platform retrieves records from the cloud database for operator verification and mission scheduling. Overall, the system operates in three stages:

\begin{framed}
{
\begin{itemize}[leftmargin=*]
    \item \textit{Stage 1: User Record Tracking and Search Event Creation.} The system collects user records (via the mobile application and cloud database), supports operator-side verification of these records, and, upon confirmation, creates a search event on the control platform.
    \item \textit{Stage 2: Search Mission Scheduling.} The system retrieves verified user records (via the control platform and cloud database), generates a search route for UAV-dispatch missions (using mode-specific strategies), and confirms the UAV-dispatch mission based on the operator's decision. 
    \item \textit{Stage 3: UAV Mission Execution and Recording.} The system dispatches the UAV to execute the search mission, during which the UAV collects updated mobile application records and captures videos via a downward-facing camera.
\end{itemize}
\par\unskip
}
\end{framed}

The detailed system design in three stages is given below.

\subsection{Stage 1: User Record Tracking and Search Event Creation}
\label{subsec:rescue_event}
In \textit{Stage~1}, the mobile application uploads user records to the cloud database. On the control platform, the operator reviews the synchronized user records and verifies their reliability and urgency. Then the operator decides whether to create a search event, which becomes the foundation for subsequent mission scheduling and UAV dispatch.

\subsubsection{User records from three service modes}
To capture different urgency levels and usage scenarios, the mobile application is designed to support three service modes:

\begin{itemize}[leftmargin=*]
    \item \textbf{Mode 1: Event Booking.} The user selects this mode to input a planned hiking route in advance. In this mode, the system can estimate the expected end time of the route and later check whether the trip is overdue. 
    \item \textbf{Mode 2: Quick Start.} The user selects this mode to start a new hiking route immediately. In this mode, the mobile application continuously records the Global Positioning System (GPS) location and time throughout the hiking process.
    \item \textbf{Mode 3: SOS.} The user selects this mode to send ``help'' signals to the ground station. In this mode, the mobile application records the current location and the request time.
\end{itemize}

When mobile connectivity is available, the mobile application uploads the records generated in any of the three modes to the cloud database. These records form a unified data stream that may later trigger search events, but are not treated as search events by default.

\subsubsection{Search event creation}
To create search events from the three service modes, the control platform first raises a notice for operator verification when specific conditions are met: a \textbf{Mode~1} trip exceeds its expected end time, \textbf{Mode~2} receives no new GPS sample for a configured interval, or the user sends an SOS request in \textbf{Mode~3}. This design ensures that search events are only created after human-in-the-loop validation, improving both reliability and resource efficiency.

Upon receiving a notice, the operator decides whether to create a search event based on the user's records. The operator first attempts to contact the user using the stored contact details. If the user cannot be reached or their safety remains uncertain, the operator contacts the emergency contact and considers the available route and location records. If no search is deemed necessary, the operator closes the notice. Otherwise, the operator confirms that a search should be initiated, and the platform creates a search event accordingly.

For \textit{Stage~2} route generation, each search event retains the relevant data: the planned route, GPS history, or SOS position, while all original user records remain accessible for review. The resulting search events enter the platform's pending queue, where they await operator dispatch and route confirmation.
\subsection{Stage 2: Search Mission Scheduling}
\label{subsec:mission_dispatch}
\textit{Stage~2} supports the operator in selecting search events and preparing UAV-dispatch missions. In this stage, verified user records are transformed into concrete UAV flight routes, with mode-specific route generation that leverages the available information from the mobile application. The operator reviews pending search events based on each user's planned route, GPS history, or SOS request and selects an event to handle. The control platform then presents a candidate flight route for the selected event on a map. The route consists of positions, called waypoints, that the UAV must visit in sequence. Each service mode contributes waypoints as follows:
\begin{itemize}[leftmargin=*]
    \item For \textbf{Mode~1}, the ground station generates waypoints from the  planned hiking route, preserving the original route plan.
    \item For \textbf{Mode~2}, it generates waypoints from the GPS histories, ordered from the earliest sample to the latest. 
    Repeated records are retained to preserve temporal information.
    \item For \textbf{Mode~3}, it generates an expanding-square spiral centered on the SOS position. The UAV flies to that position and follows the spiral route to explore the surrounding area.
\end{itemize}

After reviewing the map and associated user records, the operator confirms whether to dispatch the mission. Upon confirmation, the platform transmits the mission parameters and waypoints to the UAV. Undispatched events remain in the pending queue for later consideration. Once the UAV accepts a mission, the system enters \textit{Stage~3}.
\subsection{Stage 3: UAV Mission Execution and Recording}
\label{subsec:UAV_execution}
\textit{Stage~3} involves collecting and reporting data from the UAV during the mission to support the ongoing search event. Specifically, the UAV must upload status reports, relay collected phone records, and store flight video onboard for later review. This integrated data collection enables both real-time monitoring and post-flight analysis.

\subsubsection{Mission execution and status reporting}
Onboard software checks the mission configuration and waypoints, converts geographic coordinates into navigation goals, and executes these goals in the specified order. The UAV hovers for a configured interval at each waypoint before proceeding to the next one. When return to launch (RTL) is enabled, the software appends the launch position to the end of the route and requests landing after the final hover.

Throughout execution, the UAV streams operational telemetry over its Wi-Fi link to the ground station, covering mission acceptance, en-route transit, waypoint arrivals, and landing requests. Waypoint progression is managed onboard: the system advances to the next waypoint after confirming arrival at the current waypoint and after the hover interval has elapsed.

\begin{figure*}[!t]
    \centering
    \captionsetup[subfloat]{justification=centering}
    \subfloat[User service pages in the mobile application]{\includegraphics[width=0.47\textwidth]{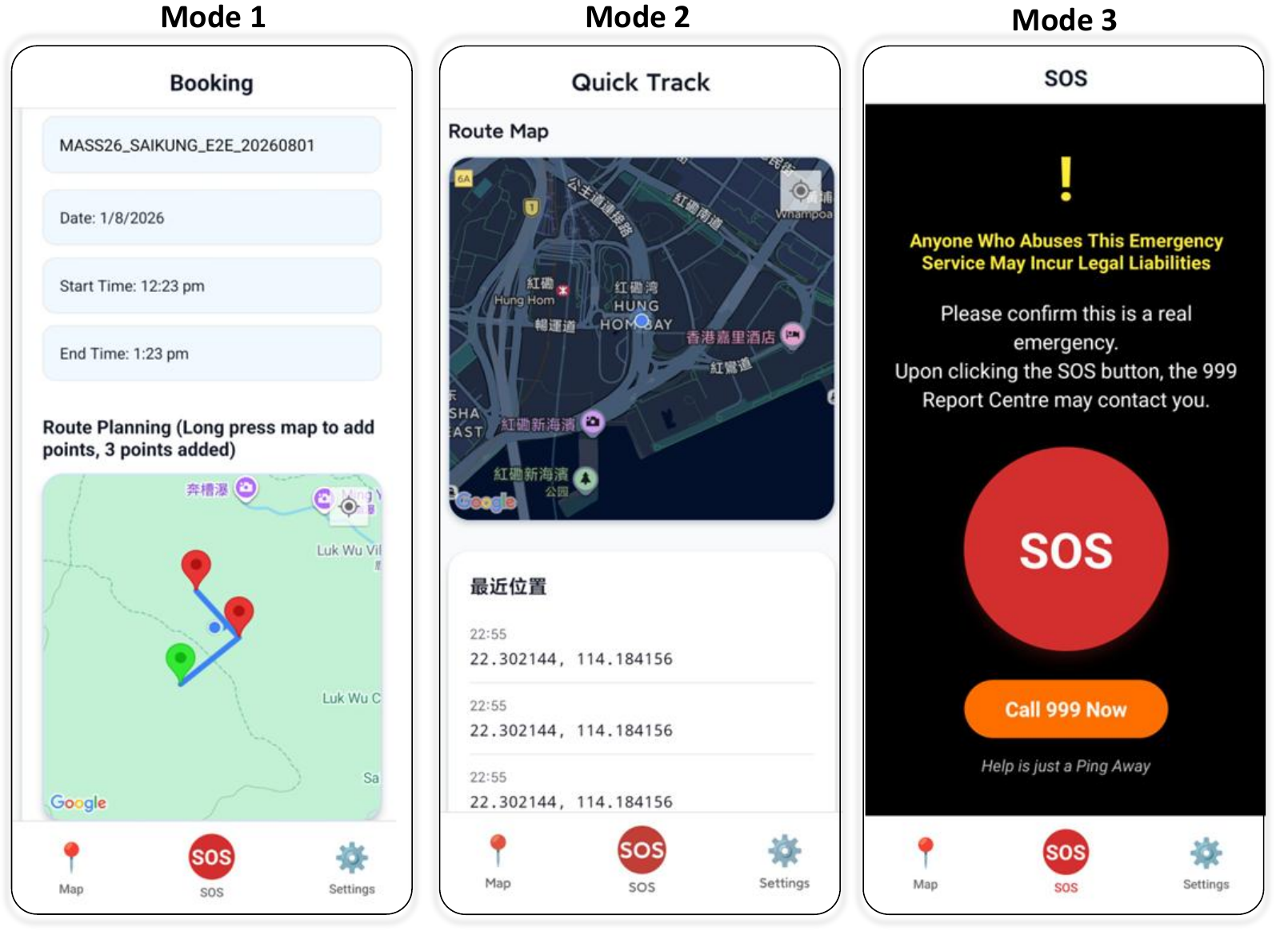}\label{fig:mapp}}\hfil
    \subfloat[Control platform used by the operator]{\includegraphics[width=0.49\textwidth]{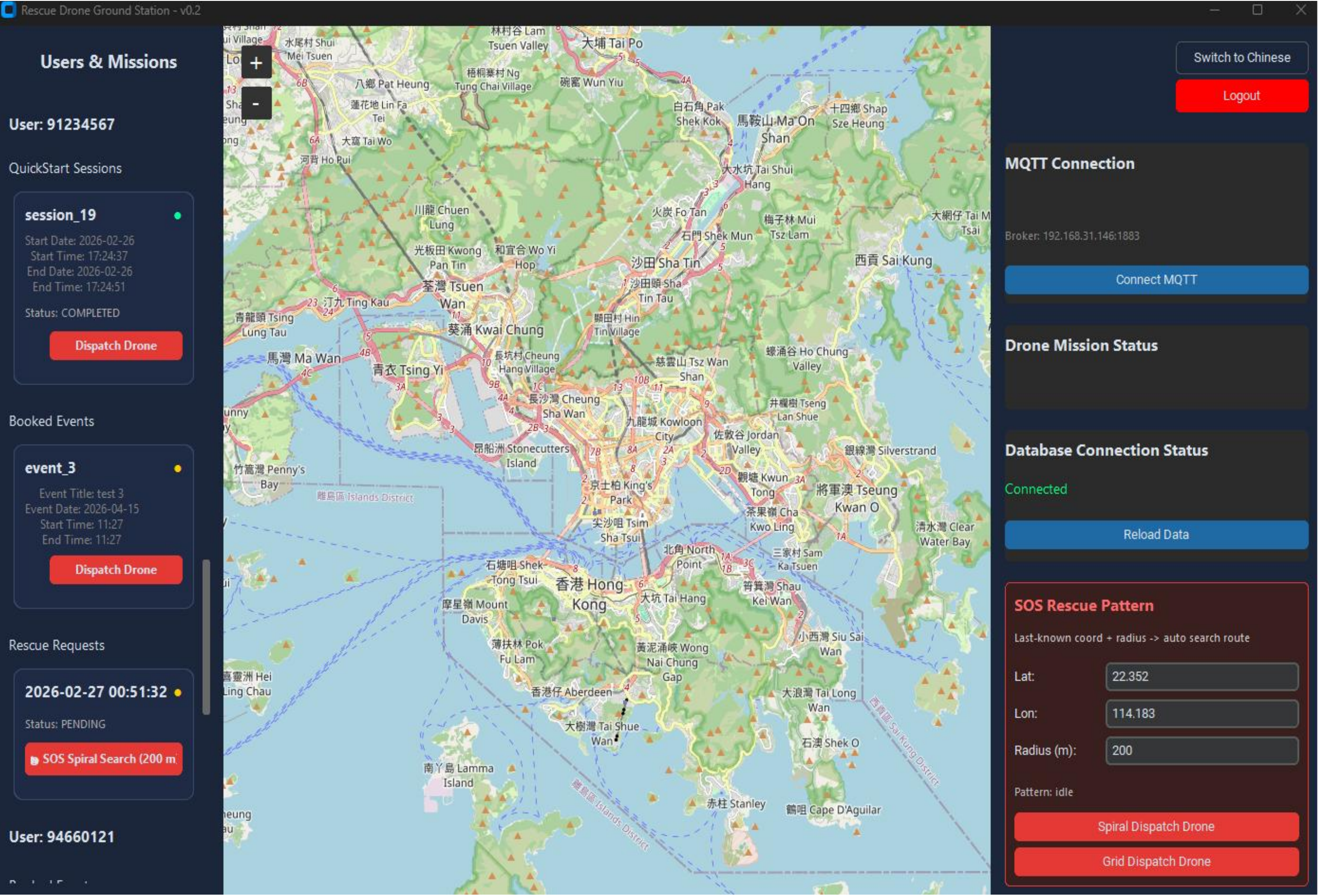}\label{fig:station}}
    \caption{Interfaces of the mobile application and the control platform. }
    \label{fig:interfaces}
\end{figure*}

\subsubsection{Phone-record collection and relay}
Once the UAV reaches a waypoint, the user can connect to its Wi-Fi hotspot to upload SOS data and updated GPS coordinates via the mobile application. Onboard software validates and stores these records together with the corresponding user and mission metadata. After the UAV completes all waypoints and requests landing, the onboard software forwards the stored records to the ground station. The control platform saves each record and returns an acknowledgment. The UAV marks the transfer as complete upon receiving the acknowledgment; otherwise, it retransmits the same record until confirmation arrives. This reliable relay mechanism ensures that critical user data is not lost, even under intermittent connectivity.

\subsubsection{Flight video recording}
The downward-facing camera records the ground along the flight route and saves the video on the onboard computer. After the flight, the operator can review people and their surroundings in the video and consider the received phone records when planning further search actions. The combination of visual evidence and user-triggered records enhances situational awareness and supports more informed decision-making in mountain search missions.

\section{System Implementation and Evaluation}
\label{sec:implementation_results}

This section describes the implementation and experimental evaluation of the whole system, including the mobile application, cloud database, control platform, and UAV prototype\footnote{Refer to our GitHub repository at \url{https://github.com/Mikeahhh/user-triggered-uav-dispatching-system} for the implementation details.}. 
The evaluations comprise search mission simulations, local software verification, and outdoor flight trials. 

\subsection{System Implementation}
\label{subsec:prototype_implementation}

\subsubsection{The mobile application and cloud database}
The mobile application is implemented using React Native and TypeScript~\cite{react_native}. \cref{fig:interfaces}(\subref*{fig:mapp}) shows the implemented interface pages of three service modes designed in \cref{subsec:rescue_event}, where ``Quick Track'' means the Quick Start service and ``Rescue'' denotes search operations. 
In \textbf{Mode~1}, the expected end time is stored as a timestamp in milliseconds, enabling consistency checks against the expected trip end time. 
In \textbf{Mode~2}, GPS coordinates and sampling times are first stored in a local SQLite database and then uploaded, preserving the full GPS history for later search planning. The cloud database is deployed on Firebase Realtime Database~\cite{firebase_realtime_database} and maintains user records, including planned routes (in \textbf{Mode~1}), GPS histories (in \textbf{Mode~2}), and SOS requests (in \textbf{Mode~3}).

\subsubsection{Control platform}
The platform is implemented in Python. Particularly, CustomTkinter and TkinterMapView are used to provide the graphical user interface and map rendering. As shown in \Cref{fig:interfaces}(\subref*{fig:station}), the implemented interface achieves map visualization, user records, mission controls, and UAV status monitoring in a single workspace, thus allowing the operator to inspect user records, select an event, prepare its route, and confirm UAV dispatch. Once an event is selected, the platform displays the corresponding waypoints. Then the operator reviews the route and confirms dispatch and the platform issues the search mission to the UAV. During execution, the platform tracks UAV status and mission progress, giving the operator an overview of both the search plan and the UAV behavior.

\subsubsection{The UAV prototype}
\begin{figure}[!t]
    \centering
    \includegraphics[width=0.95\columnwidth]{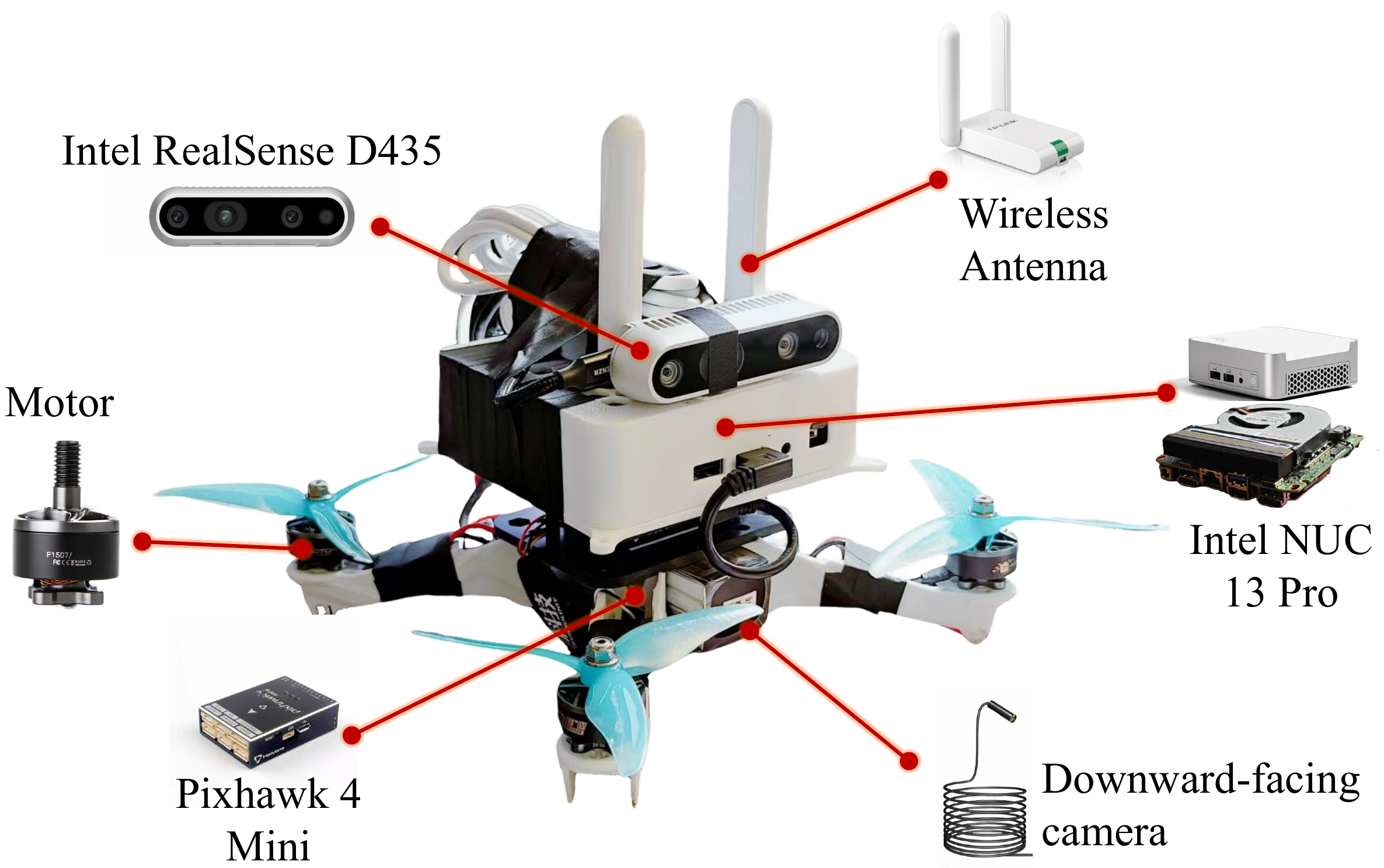}
    \caption{The UAV prototype. }
    \label{fig:UAV_hardware}
\end{figure}

As shown in \cref{fig:UAV_hardware}, the UAV prototype integrates an Intel NUC~13 Pro onboard computer for mission processing and recording, a Pixhawk~4 Mini for low-level flight control, and Wi-Fi modules for communication between the UAV and ground station.
Sensing capabilities are provided by an Intel RealSense D435 depth camera for obstacle avoidance and path planning, complemented by a downward-facing camera for ground imaging during search missions. Python software on the onboard computer receives missions from the ground station and manages waypoint execution order. C++ software converts geographic waypoints into navigation goals compatible with the flight stack.
These programs operate with Fast-Drone-250's Robot Operating System (ROS) stack~\cite{karjalainen2026towards}, with \texttt{px4ctrl}\footnote{\texttt{px4ctrl} is an onboard flight-control software module. } 
generating flight-control commands for the Pixhawk~4 Mini. Phones upload records through the UAV hotspot; onboard software stores these records and later forwards them to the ground station. Video is recorded on the onboard computer for post-mission review. For safety and manual override, a RadioLink AT9S Pro transmitter and R12DSM receiver provide remote control, with SBus linking the receiver to the flight controller. This combination enables both autonomous operation within the proposed system and manual intervention when required.

\subsection{Simulation of Search Missions}
\label{subsec:mission_simulations}
To assess the behavior of the proposed modes under realistic terrain conditions, 
MATLAB simulations were conducted to evaluate the full flight cycle of takeoff, execution, return, and landing for all three modes. 

\subsubsection{Simulation scenario}
The simulation scenario is based on ``Sai Kung, Hong Kong'', with the launch site located near Pak Tam Chung at $22.402^\circ$N, $114.322^\circ$E. Terrain elevations are taken from the Mapzen/Tilezen Skadi tile N22E114, a $3601\times3601$ grid with one-arc-second spacing~\cite{skadi_terrain}. Bilinear interpolation provides a 25~m plotting grid for visualization, without increasing the underlying source resolution. Elevations are referenced to the Earth Gravitational Model 1996 (EGM96) geoid. Candidate target locations were sampled uniformly between $1.5$ and $4$km east and $0.5$ and $4$km north of the launch, requiring a launch-to-target distance of $2$ to $4.5$km and an elevation between $100$ and $500$m. Land checks required elevations of at least $40$m on a $17\times17$ grid extending $240$m in each direction around the candidate target, and minimum clearance of $2$m along all mission and return segments sampled at intervals no greater than $5$m. 
The sixth candidate satisfied all constraints and was selected as the common target: $22.40782761^\circ$N, $114.35687745^\circ$E, at approximately $177.44$m elevation. This target was used for all three modes to enable direct comparison of mission behavior.

\begin{figure*}[!t]
    \centering
    \includegraphics[width=0.86\textwidth]{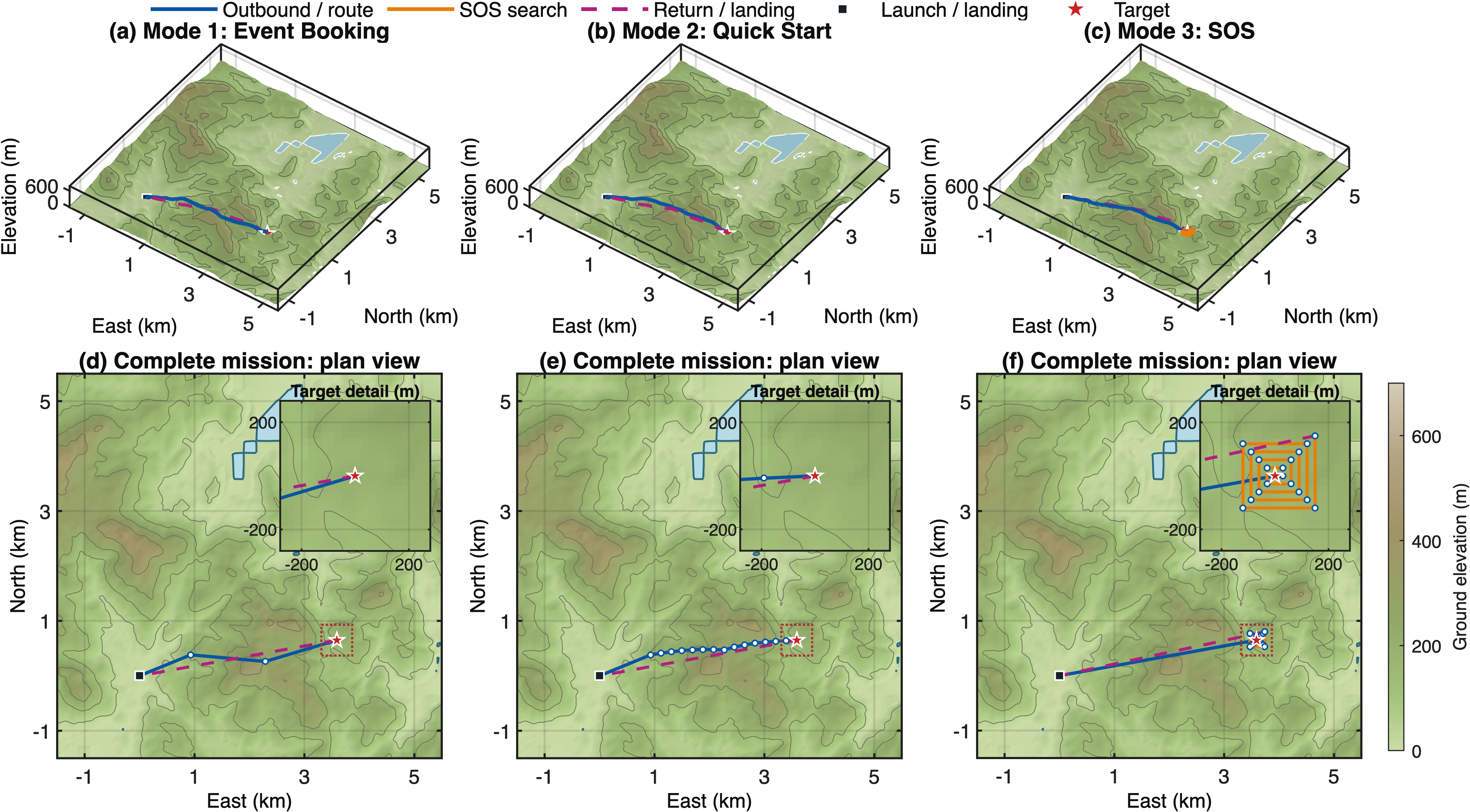}
    \caption{Search mission simulations: (a, d) \textbf{Mode~1}, (b, e) \textbf{Mode~2}, and (c, f) \textbf{Mode~3}. Top: three-dimensional paths; bottom: search plan views. }
    \label{fig:terrain_missions}
\end{figure*}

\subsubsection{Routes for the three modes}
The planned route and GPS history were constructed for this simulation to illustrate how the three modes exploit different types of user records:

{
\begin{itemize}[leftmargin=*]
    \item For \textbf{Mode~1}, the UAV sequentially traversed three waypoints placed at $27\%$, $63\%$, and $100\%$ of the launch-to-target baseline, with respective lateral offsets of $210$m left, $150$m right, and $0$m from the baseline, and then returned directly to the launch site.
    \item For \textbf{Mode~2}, fifteen samples were spaced equally along the launch-to-target direction between the first and last Mode~1 waypoints. A smooth half-sine detour with a maximum lateral offset of $210$m was superimposed on the piecewise-linear Mode~1 route, emulating a realistic hiking path. Timestamps were generated assuming a walking speed of $1$m/s, and all samples were available before dispatch. The UAV followed this stored GPS history in chronological order, then returned to the launch site.
    \item For \textbf{Mode~3}, the UAV executed a counterclockwise expanding-square spiral starting $30$m east of the centre, with leg length incremented by $30$m every two legs. Waypoint generation terminated at the first endpoint at least $200$m from the centre, yielding 18 spiral waypoints plus the centre point. All waypoints were visited before the UAV returned to launch.
\end{itemize}
}

\subsubsection{Flight settings}
For all modes, the UAV performed a vertical takeoff and then maintained a terrain-following altitude of $80$m above ground level throughout transit, route execution, and return. Horizontal and vertical speeds were limited to $15$ and $3$m/s, respectively. The UAV hovered for $5$~s at each mission waypoint, returned directly to the launch site, hovered for an additional $5$~s, and then executed a vertical landing. Route segments were sampled in the horizontal plane at intervals no greater than $5$m. The duration of each step was set to the maximum of the horizontal distance divided by $15$m/s and the absolute elevation change divided by $3$m/s, ensuring that both speed limits were respected. Timing began at takeoff and did not include operator verification, dispatch operations, or communication delays, isolating the flight behavior of the system.

\subsubsection{Simulation results}
\Cref{fig:terrain_missions} presents the three-dimensional mission paths together with plan views and detailed insets around the target. The main axes share the launch origin and span $-1.5$ to $5.5$km east and north. Blue segments represent outbound flight and user-route traversal, orange segments represent the spiral search pattern, and dashed magenta segments represent return and landing. In \textbf{Mode~1}, the UAV followed the planned route through its three waypoints and then returned to the launch site. In \textbf{Mode~2}, the UAV reconstructed and followed the user’s GPS history in time order before returning. In \textbf{Mode~3}, the UAV reached the target area, completed the expanding-square spiral search, and then flew back to the launch site. In all modes, waypoints were visited in the intended order and the UAV landed at the launch site, confirming that the mission logic and trajectory generation function correctly across the three user-triggered modes. The takeoff-to-landing times were $715.7$, $764.1$, and $1008.5$~s for Mode~1, Mode~2, and Mode~3, respectively.

\begin{figure*}[!t]
    \centering
    \includegraphics[width=0.86\textwidth]{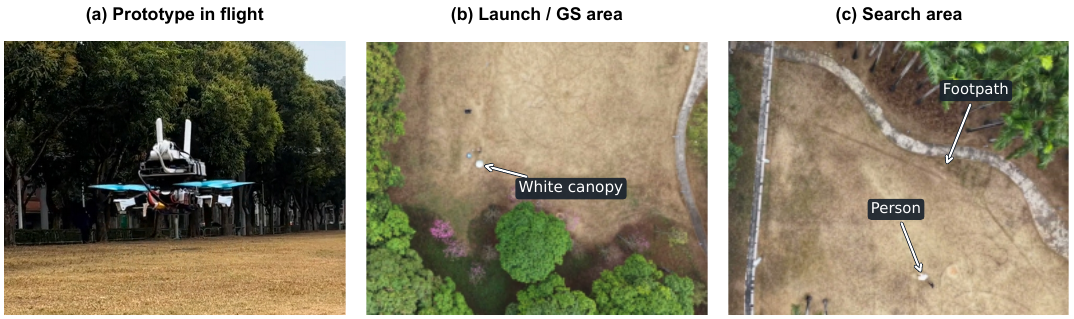}
    \caption{Outdoor flight and video recording: (a) the UAV during flight; (b) the launch and ground station area; and (c) the search area. } 
    \label{fig:outdoor_tests}
\end{figure*}

\subsection{Local Software Verification}
\label{subsec:local_software_verification}

To verify the correctness of the local software components without relying on full hardware deployment, we conducted tests using simulated user data. These tests focused on timeout decisions and the reception and relay of phone records, covering both timeout notifications and the reliable transfer of stored records from the UAV software to the ground station.

\subsubsection{Location-update timeout decisions}
We constructed synthetic GPS positions and sampling times to test whether \textbf{Mode~2} correctly detects a timeout relative to the latest valid sample. The program compares the elapsed time between the moment of the check and the timestamp of the most recent sample against a configured threshold. Using a $1$~s threshold, we evaluated elapsed times of $0.999$, $1.000$, and $1.001$~s, representing values just below, exactly at, and just above the boundary condition. Timeouts were triggered at $1.000$~s and $1.001$~s but not at $0.999$~s, confirming that location updates time out precisely when the elapsed interval reaches the configured threshold and not before.

\subsubsection{Phone-record reception and relay}
We ran the UAV receiver and relay modules alongside the ground station receiver on a local computer to test record handling end to end. A synthetic phone record contained SOS information, a GPS position, and a user identifier. An emulated phone upload exercised the full path from the mobile application, through the UAV software, to the ground station. After the UAV receiver stored the record, a simulated waypoint completion and landing request triggered data relay to the ground station. The ground station then archived the record and returned an acknowledgment, prompting the UAV software to finalize the transfer and mark the record as delivered. Cross-verification confirmed that records stored at both endpoints retained identical SOS coordinates, GPS logs, user identifiers, and mission associations. Additional scenarios investigated acknowledgment loss and interference between missions. In the absence of acknowledgment, records were repeatedly retransmitted until confirmation was received, demonstrating robustness against communication loss. Foreign mission completion messages were correctly ignored and did not trigger data relay, ensuring that records are associated only with their intended missions.
\subsection{Outdoor UAV Tests}
\label{subsec:outdoor_UAV_tests}

To demonstrate the integrated operation of the proposed system in a realistic environment, we conducted outdoor tests that simulated a complete search event, from mission preparation at the ground station to visual inspection in the search area. \Cref{fig:outdoor_tests}(a) shows the UAV prototype in flight. The ground station is located near the white canopy at the launch site, and the UAV departs from this position toward the designated search area, as illustrated in \Cref{fig:outdoor_tests}(b). \Cref{fig:outdoor_tests}(c) shows the search area, with a person standing next to a footpath and surrounding vegetation. These spatial relationships in the recorded images help the operator assess whether the area warrants further inspection. The saved images support post-flight review of both personnel and the surrounding environment during subsequent searches. Together, these outdoor tests demonstrate how the UAV’s flight and recording functions cooperate in a search mission: the UAV autonomously reaches the designated area, captures imagery of the scene, and provides visual information that the operator can use to determine whether additional investigation or assistance is required. This illustrates the UAV’s role in the proposed user-triggered system as a rapid and precise means of reaching mountain search areas and collecting actionable visual data.


\section{Conclusion}
\label{sec:conclusion}

We presented a user-triggered UAV dispatching system for precise and fast mountain search. The system links planned hiking routes, GPS location histories, and SOS requests from hikers’ smartphones to UAV dispatch. An operator verifies each alert or anomaly, creates a search event that aggregates the relevant records, and reviews the automatically generated flight route before confirming dispatch. The UAV then executes the approved mission, follows waypoints, reports its status, relays phone records when available, and records downward-facing imagery for later inspection. The prototype implementation was evaluated through software tests, which confirmed correct timeout handling and record transfer according to the specified rules, and through outdoor field trials, which demonstrated autonomous flight and ground image collection suitable for operator review. These results suggest that integrating hiker-side information, cloud-based event management, and UAV search provides a practical basis for faster and more precise mountain search and rescue missions.

Future work will evaluate the system through end-to-end field tests in mountain environments with intermittent connectivity, focusing on data transmission success rates and response latency. We will also measure the energy consumption of continuous GPS recording in Mode~2, together with the UAV's effective flight endurance and coverage radius during search missions. These evaluations will inform route planning that accounts for battery reserves, communication availability, and uncertainty in user locations. We will further extend the ground station's mission scheduling to support multi-UAV cooperative searches through search-area allocation and the sharing of mission progress and user records.

\section{Acknowledgement}
This work was supported in part by the Research Grants Council of the Hong Kong Special Administrative Region, China, under Project UGC/FDS16/E15/24, in part by the Hong Kong Metropolitan University Research Grant under Project PFDS/2025/34, in part by the UGC Research Matching Grant Scheme under Project 2024/3003 and Gekko Lab, and in part by the National Natural Science Foundation of China under Grant 62402391.

\balance
\renewcommand{\IEEEbibitemsep}{0pt plus 0.5pt}
\bibliographystyle{IEEEtran}
\bibliography{ref}

\end{document}